\documentclass[10pt, conference, letterpaper]{IEEEtran}
\IEEEoverridecommandlockouts
\usepackage{cite}
\usepackage{amsmath,amssymb,amsfonts}
\usepackage{algorithm}
\usepackage[noend]{algpseudocode}
\usepackage{subcaption}
\usepackage{booktabs}
\usepackage{graphicx}
\usepackage{comment}
\usepackage{enumitem}
\usepackage{textcomp}
\usepackage{multirow}
\usepackage{xcolor}
\usepackage{xspace}

\usepackage{pifont}
\newcommand{\chen}[1]{}
\newcommand{\phm}[1]{\vspace{.4em} \noindent\textbf{#1}\hspace{.5em}} 

\def\BibTeX{{\rm B\kern-.05em{\sc i\kern-.025em b}\kern-.08em
    T\kern-.1667em\lower.7ex\hbox{E}\kern-.125emX}}

\begin{document}
\bstctlcite{IEEEexample:BSTcontrol}

\title{\textsc{Pulse}: Training Acceleration for Large Diffusion Models with Automatic Pipeline Parallelism
}
\newcommand{\name}{\textsc{Pulse}\xspace}

\author{\IEEEauthorblockN{
Boran Sun\textsuperscript{1*},
Guoyong Jiang\textsuperscript{2,5,*},
Lin Zhang\textsuperscript{1\dag},
Chen Chen\textsuperscript{3},
Yuechen Tao\textsuperscript{5},
Zhishu Che\textsuperscript{4},
Jieling Yu\textsuperscript{1},\\
Shan Chang\textsuperscript{4},
Huaxi Gu\textsuperscript{2},
Fangming Liu\textsuperscript{5},
Bo Li\textsuperscript{1\dag},
}
\IEEEauthorblockA{
\textsuperscript{1}The Hong Kong University of Science and Technology,
\textsuperscript{2}Xidian University,
\textsuperscript{3}Shanghai Jiao Tong University,\\
\textsuperscript{4}Donghua University,
\textsuperscript{5}Pengcheng Laboratory\\
Emails: \{bsunak, lzhangbv, ytaoaf, jyucm\}@connect.ust.hk, gyjiang@stu.xidian.edu.cn,\\
chen-chen@sjtu.edu.cn, che\_zs@mail.dhu.edu.cn, changshan@dhu.edu.cn,\\
hxgu@xidian.edu.cn, fangminghk@gmail.com, bli@cse.ust.hk
}
\thanks{*Equal contribution.}
\thanks{\dag Corresponding authors.}
}

\maketitle

\begin{abstract}

Diffusion models are now a dominant approach for high-fidelity image and video generation, yet scaling their training across GPU clusters remains challenging. Unlike transformer-only architectures, diffusion backbones commonly adopt UNet-style encoder-decoder structures with heterogeneous layers and long-range skip connections. Under conventional pipeline parallelism, these non-local dependencies force large skip activations and their gradients to traverse multiple pipeline boundaries, making peer-to-peer (P2P) communication a dominant bottleneck and substantially reducing pipeline efficiency. In this paper, we present \name, an automatic pipeline-parallel training strategy that makes \emph{skip locality} a first-class optimization objective. \name eliminates skip-induced communication by collocating skip-connected encoder--decoder layers on the same device and caching skip activations locally for later use in backpropagation. To realize this placement while maintaining high pipeline utilization, \name co-designs: (1) a skip-aware dynamic-programming partitioner that balances heterogeneous stage workloads under symmetric collocation constraints, (2) an ILP-based schedule synthesizer that generates bubble-efficient wave schedules for the resulting stage-to-device mapping, and (3) a hybrid parallelism tuner that selects pipeline/data-parallel degrees and microbatch sizes under memory and network constraints. Our extensive experiments show that the volume of communication can be reduced by 89\%, and the training throughput can be increased by up to 2.3$\times$ on communication-bound hardware, compared with state-of-the-art parallelism strategies.

\end{abstract}

\begin{IEEEkeywords}
    Distributed Deep Learning; Diffusion Models; Communication Optimization; Automatic Pipeline Parallelism
\end{IEEEkeywords}

\section{Introduction}

\begin{figure}[htpb]
    \centering
    \includegraphics[width=\linewidth]{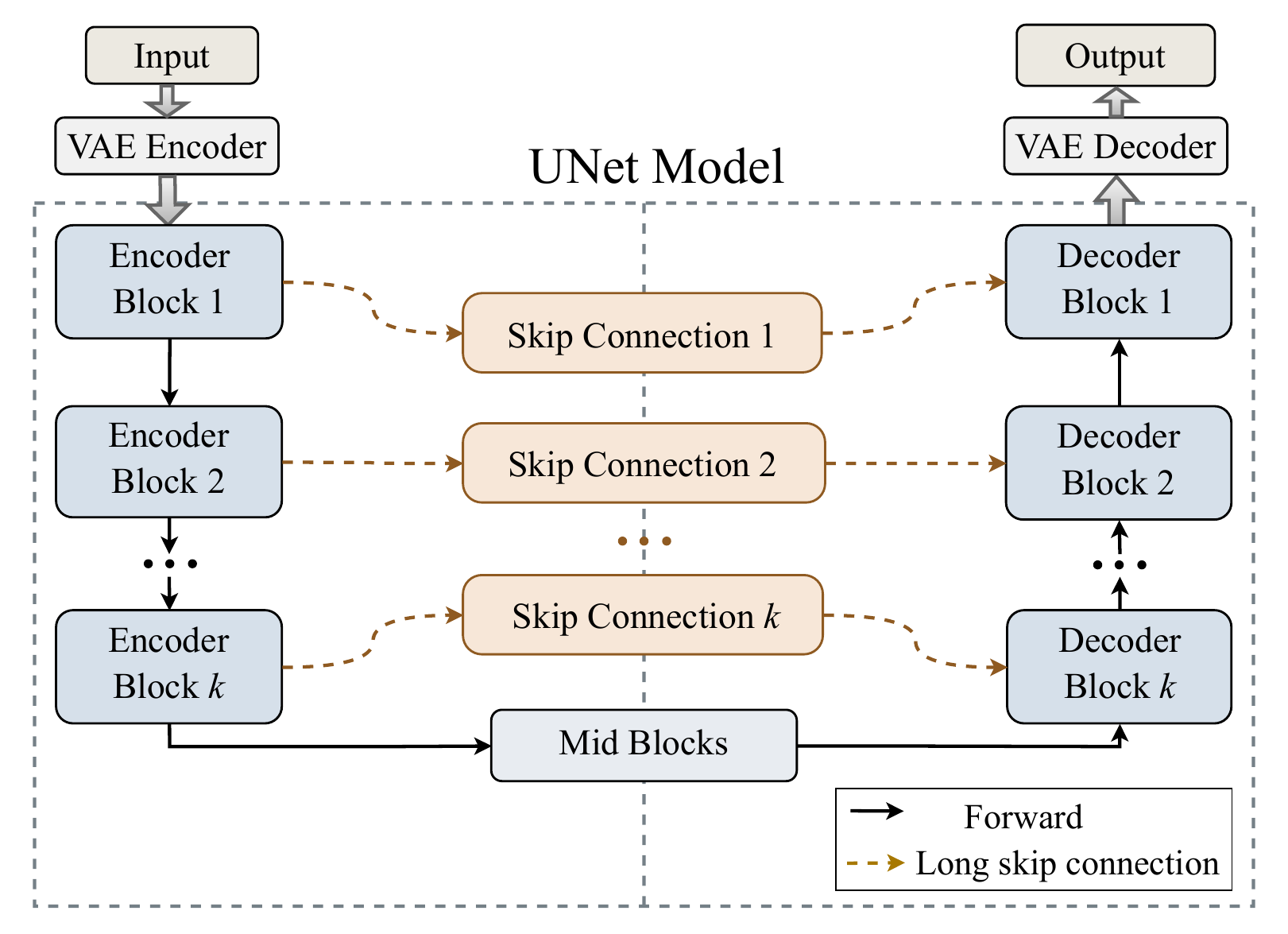}
    \caption{The latent diffusion architecture used by diffusion models. The model architecture contains symmetric skip connections between the encoder and decoder blocks.}
    \label{fig:overview_arch}
\end{figure}

Diffusion models have become a foundation of modern generative AI for synthesizing high-resolution images and videos~\cite{rombach2022high,ho2020denoising,saharia2022photorealistic,ramesh2022hierarchical,balaji2022ediff}.  They are widely deployed in text-to-image generation~\cite{rombach2022high,peebles2023scalable,li2024hunyuan,bao2023all}, image editing~\cite{avrahami2022blended}, and video synthesis~\cite{blattmann2023stable}. As illustrated in Fig.~\ref{fig:overview_arch}, many diffusion backbones follow a UNet-style \emph{encoder--decoder} architecture with \emph{symmetric skip connections} that preserve spatial detail.  In a typical latent diffusion pipeline, a variational autoencoder (VAE) encodes an image into a latent representation, and a UNet-like denoiser iteratively predicts and removes noise from the latent, optionally conditioned on text or other modalities.  Recent diffusion models continue to increase in scale (e.g., multi-billion-parameter denoisers), making distributed training on commodity accelerators increasingly necessary~\cite{esser2024scaling,labs2025flux1kontextflowmatching}.

A widely adopted distributed strategy for diffusion training is data parallelism (DP), where each device processes a subset of the batch and gradients are synchronized across replicas. Memory-optimization techniques such as ZeRO~\cite{rajbhandari2020zero} and Fully Sharded Data Parallel (FSDP)~\cite{zhao2023pytorch} reduce per-device memory usage by partitioning parameters and optimizer states. However, these approaches introduce frequent collective communications (e.g., AllGather/ReduceScatter), which can become a major bottleneck on bandwidth-constrained hardware and in multi-node settings. As shown in Fig.~\ref{fig:zero3_comm}, communication can consume a substantial fraction of iteration time even on high-bandwidth interconnects.

Pipeline parallelism (PP)~\cite{narayanan2021efficient,huang2019gpipe} is a complementary approach that partitions the model into stages across devices and pipelines microbatches through forward and backward passes. PP has been highly effective for large language models with near-sequential dataflow, but diffusion models pose a key challenge: UNet-style skip connections induce long-range dependencies that violate the sequential assumptions implicit in common pipeline schedules (e.g., 1F1B). With a naive sequential partition, skip activations produced by early encoder stages must be transmitted across multiple stage boundaries before reaching their corresponding decoder stages, and the associated gradients must traverse back during backpropagation.  Empirically, this skip-induced traffic dominates P2P communication volume in pipeline-parallel diffusion training (Fig.~\ref{fig:comm_breakdown}), severely limiting scalability and often negating the expected benefits of pipelining.

This paper presents \name, an automatic pipeline-parallel training system tailored to diffusion models with long-range skip connections. Our key insight is that the dominant communication overhead can be removed by enforcing \emph{skip locality}: collocating skip-connected encoder-decoder layers on the same device, and treating skip tensors as local buffers that are stored during the forward pass and reused during the backward pass.  This eliminates skip-induced inter-device transfers, reducing both bandwidth pressure and the lifetime of cached skip activations. At the same time, skip-aware collocation introduces new algorithmic constraints that existing auto-parallelism systems are not designed to handle. \emph{First}, diffusion models exhibit substantial layer heterogeneity (e.g., varying resolutions and operator types), making naive block-wise partitioning prone to severe stage imbalance and pipeline bubbles. \emph{Second}, collocation constraints restrict feasible placements and invalidate standard sequential scheduling assumptions, requiring schedules that remain correct and efficient under non-trivial stage-to-device mappings. \emph{Finally}, achieving high end-to-end throughput in multi-node environments requires jointly selecting pipeline and data-parallel degrees and microbatch sizes under memory and network constraints.

To address these challenges, we design \text{\name} with three key components:
\begin{itemize}[left=0em, itemsep=2pt, topsep=2pt]
    \item \textbf{Skip-aware partitioning.} We introduce a dynamic-programming partitioner that extends linear partitioning to a bidirectional formulation, explicitly enforcing symmetric encoder-decoder collocation while balancing heterogeneous per-stage runtimes.
    \item \textbf{Constraint-aware scheduling.} We formulate pipeline scheduling under collocation constraints and use a synthesizer based on integer linear programming (ILP) to recover efficient execution patterns. This scheduler enables efficient forward-backward pass interleaving, minimizing training steps even when skip connections disrupt the canonical layer sequence.
    \item \textbf{Hybrid parallelism optimization.} We design a hybrid parallelism tuner that systematically explores valid combinations of pipeline and data parallelism degrees. Guided by profiled memory consumption and per-layer computational costs, this tuner identifies configurations that maximize throughput while respecting hardware resource constraints.
\end{itemize}

We evaluate \name\ on three representative diffusion backbones --- UViT~\cite{bao2023all}, Stable Diffusion v2~\cite{rombach2022high}, and Hunyuan-DiT~\cite{li2024hunyuan} --- across two clusters: a 2-node NVIDIA V100 cluster (16 GPUs) and an 8-node Ascend 910A cluster (64 NPUs). Our extensive array of experimental results have shown convincing evidence that \name\ outperforms baselines including DeepSpeed ZeRO-2, Megatron 1F1B, and Hanayo pipeline parallelism. On the Ascend 910A cluster, \name\ increases training throughput by up to \textbf{2.3$\times$} and reduces communication volume by \textbf{90\%}. On the V100 GPU cluster, it achieves \textbf{16\%-125\%} higher throughput and \textbf{89\%} reduction in communication volume. This indicates that \name\ can substantially improve the performance and scalability of training diffusion models on commodity GPU clusters.

\section{Background and Motivation}
In this section, we present the background of training diffusion models and our motivations. 

\subsection{Diffusion Models with Skip Connections}

Diffusion models have emerged as the foundation of modern image generative modeling, offering high sample quality and robust training stability~\cite{rombach2022high, ho2020denoising}. In general, diffusion models use a UNet architecture~\cite{ronneberger2015u} as the backbone to predict noise in a latent space. Taking Stable Diffusion v2 (SDv2)~\cite{rombach2022high} as an example, as shown in Figure~\ref{fig:overview_arch}, the latent from VAE encoder is fed into a UNet model, which consists of encoder blocks on the left side, decoder blocks on the right side, and long skip connections between them. Each block contains Attention~\cite{vaswani2017attention} and ResNet~\cite{he2016deep} layers. With skip connections, UNet can pass high-resolution details directly from early encoder blocks to late decoder blocks, which has been shown to accelerate convergence, mitigate vanishing gradients, and improve quality for image synthesis~\cite{ronneberger2015u,li2024hunyuan}.

Modern diffusion models such as UViT~\cite{bao2023all} and Hunyuan-DiT~\cite{li2024hunyuan} retain this encoder-decoder architecture with skip connections, but replace other modules with Transformer blocks~\cite{vaswani2017attention}. Many studies have shown that these skip connections, from early Transformer blocks to late Transformer blocks, are still essential to maintain fast convergence speed and high-quality visual fidelity~\cite{bao2023all,li2024hunyuan}. For example, ablation studies in Hunyuan-DiT have shown that removing skip connections can result in a 10.3\% increase in FID and 1.8\% decrease in the CLIP score, indicating poorer image fidelity~\cite{li2024hunyuan}. In addition, recent diffusion models are increasing in parameter size and cannot fit into a single device. For example, Stable Diffusion 3.5 Large~\cite{esser2024scaling} contains 8.1B parameters, and FLUX~\cite{labs2025flux1kontextflowmatching} scales to 12B parameters. As a result, parallel strategies are required for training large diffusion models.



\subsection{Parallel Strategies for Diffusion Models}


To accommodate the growing computational demands of large diffusion models, various parallelism strategies have been developed. These methods fall broadly into data parallelism (DP), pipeline parallelism (PP), and their hybrid combinations. 

\begin{figure}[t]
  \centering
  \begin{minipage}[t]{0.49\linewidth}
    \centering
    \includegraphics[width=\linewidth]{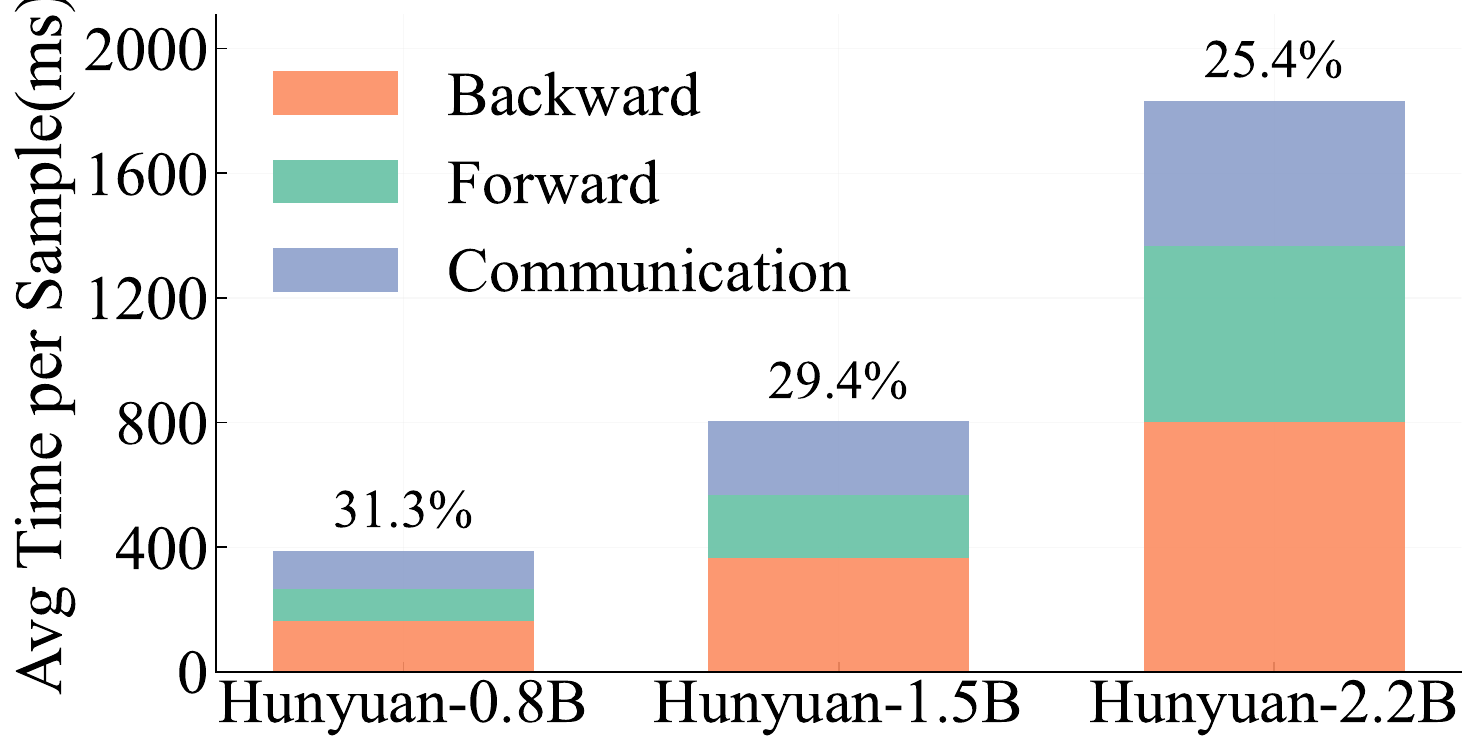}
    \caption{Time breakdown of ZeRO3 for training Hunyuan-DiT models on 8 V100 GPUs.}
    \label{fig:zero3_comm}
  \end{minipage}%
  \hfill
  \begin{minipage}[t]{0.49\linewidth}
    \centering
    \includegraphics[width=\linewidth]{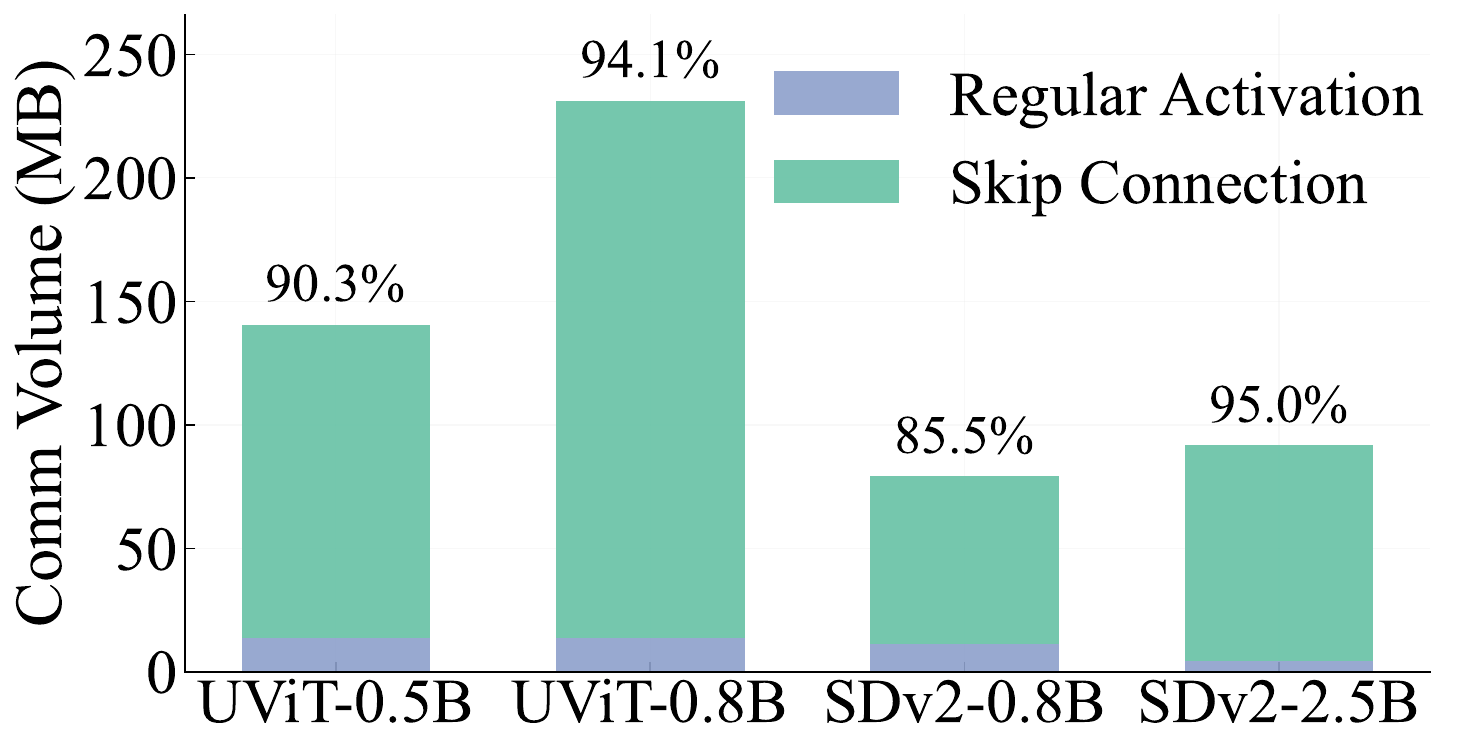}
    \caption{Communication volume breakdown of PP for models with skip connections.}
    \label{fig:comm_breakdown}
  \end{minipage}
\end{figure}

Small diffusion models are typically trained with DP, where each device holds a full model replica and processes a subset of training data. Gradients are synchronized across devices via all-reduce communications. As model size grows,  ZeRO~\cite{rajbhandari2020zero} and FSDP~\cite{zhao2023pytorch} techniques have been proposed to reduce memory usage by partitioning parameters, gradients, and optimizer states across devices. However, this comes at the cost of increased inter-device communications with frequent all-gather and reduce-scatter operations. As illustrated in Figure~\ref{fig:zero3_comm}, communication time can take up around 30\% of total training time when training Hunyuan-DiT models using ZeRO Stage3 (ZeRO-3) on 8 V100 GPUs with 300GB/s NVLink. 

On the other hand, PP is introduced to reduce memory usage by partitioning model layers into sequential stages, each assigned to a different device~\cite{huang2019gpipe,narayanan2019pipedream}. Training samples are split into micro-batches and processed in a pipelined manner to reduce device idle time. For example, Dapple~\cite{fan2021dapple} and Megatron-LM~\cite{shoeybi2019megatron} introduce 1F1B (one forward, one backward) pipeline scheduling to address the activation memory problem, while Chimera~\cite{li2021chimera} and Hanayo~\cite{liu2023hanayo} proposed bidirectional and wave-like schedule algorithms, respectively, to further reduce pipeline bubble. PP is more promising than ZeRO in addressing memory constraint under a modest communication cost, however, existing methods overlook the complexity of diffusion models with skip connections. 

\subsection{Problems of Applying PP to Diffusion Models}
\label{sec:comm_model}

Existing PP methods, such as 1F1B~\cite{shoeybi2019megatron}, typically assume sequential data flow, where data can only be passed from one stage to its previous or subsequent stage. It works well for sequential models such as decoder-only transformers in large language models, where intermediate activations are passed from one layer to its subsequent layer. However, for diffusion models with UNet-like model backbones (as shown in Figure~\ref{fig:overview_arch}), they could violate this assumption due to the existence of long-range skip connections. In diffusion models, encoder blocks need to transmit intermediate activations to decoder blocks several stages away in the forward pass, while decoder blocks send gradient activations back to the corresponding encoder blocks in the backward pass. When these skip-connected blocks are assigned to different devices, it will introduce costly inter-device communication overheads. 


\begin{figure}[!htbp]
  \centering
  \includegraphics[width=\linewidth]{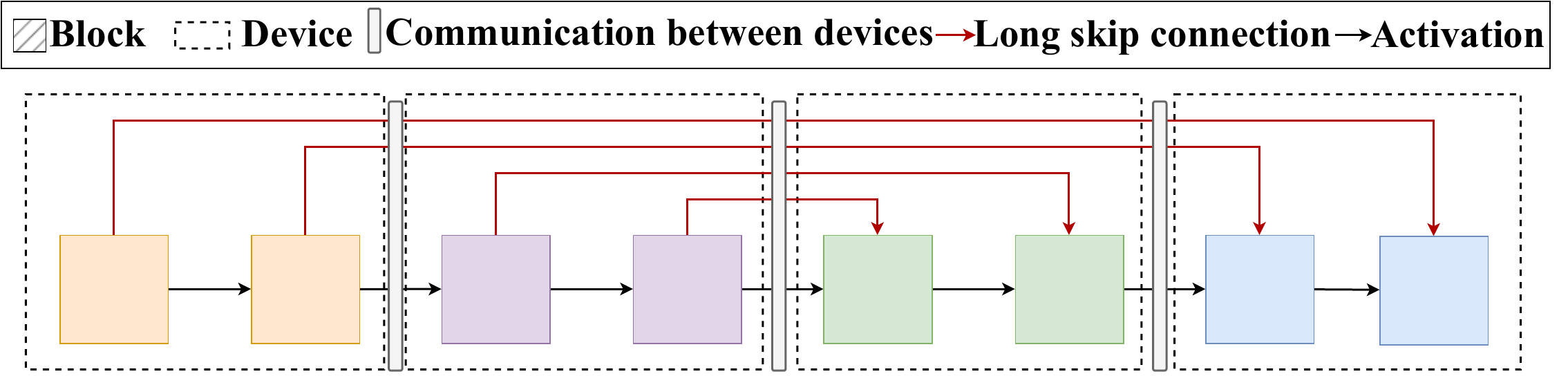}
  \caption{Illustration of inter-device communication in UNet-like models with skip connections under a 1F1B pipeline schedule. Same-color blocks denote layers assigned to the same device. Skip connections across devices require extra communication during both forward and backward passes.}
  \label{fig:1f1b_comm}
\end{figure}

For example, in Figure~\ref{fig:1f1b_comm}, we illustrate this problem of applying PP to a simplified diffusion model, which has 4 encoder blocks and 4 decoder blocks, each pair of which has a long-range skip connection. Following the common practice of 1F1B, we partition 8 blocks into 4 devices in a sequential manner (blocks with the same color are assigned to the same device). Consequently, activations passed via skip connections should be transmitted from one stage to its next stage multiple times until it arrives destination stage. These additional transfers largely increase the total communication volume. Formally, for a model with $K$ blocks distributed over $D$ devices, assuming each block produces activations of size $a$, the total communication volume can be calculated as $\left(\frac{(K+4)D}{4}-1\right)a$. As model depth or pipeline width increases, the skip-induced traffic scales linearly, leading to poor scalability and suboptimal device utilization.

To understand the overhead introduced by skip connections, we measure the communication volume in a single micro batch forward process of diffusion models under a 4-device pipeline parallel setting. As shown in Figure~\ref{fig:comm_breakdown}, skip connections account for the vast majority of inter-device communication, exceeding 85.5\%-90\% in UViT and SDv2 models and dominating bandwidth, which adversely affects the end-to-end training performance of applying PP to diffusion models, as evaluated later in Figure~\ref{fig:main_exp}. 

\subsection{Opportunities and Challenges}

To address the inefficiency caused by skip connections, we are motivated to build an automatic pipeline parallelism system to maximize training throughput for large diffusion models. However, we find that existing auto-parallelism frameworks~\cite{fan2021dapple,zheng2022alpa} require models to be expressed as a sequence of layers and fail to handle model complexity and long-range communication patterns introduced by skip connections. To solve this, our solution is to collocate skip-connected layers to the same device with two key advantages. First, it allows us to apply PP safely without breaking the sequential data flow, as only short-range connections can span device boundaries. Second, it eliminates expensive skip-connection-induced communications by storing intermediate activations on the same device for subsequent computation. 

There are three major technical challenges to realize efficient and skip-aware PP with automatic parallelism. First, diffusion models have an encoder-decoder structure with layers of different sizes and computational demands, which could cause uneven workload distributions across pipeline stages, leading to idle GPU time (i.e., pipeline bubbles). Second, skip connections between encoder and decoder layers create dependencies that disrupt standard pipeline scheduling. This requires new strategies to coordinate micro-batch execution across sequential stages. Third, combining PP and DP involves navigating trade-offs between memory usage, communication overhead, and computational efficiency. 

\section{Overview}
\begin{figure}[!htbp]
  \centering
  \includegraphics[width=0.99\linewidth]{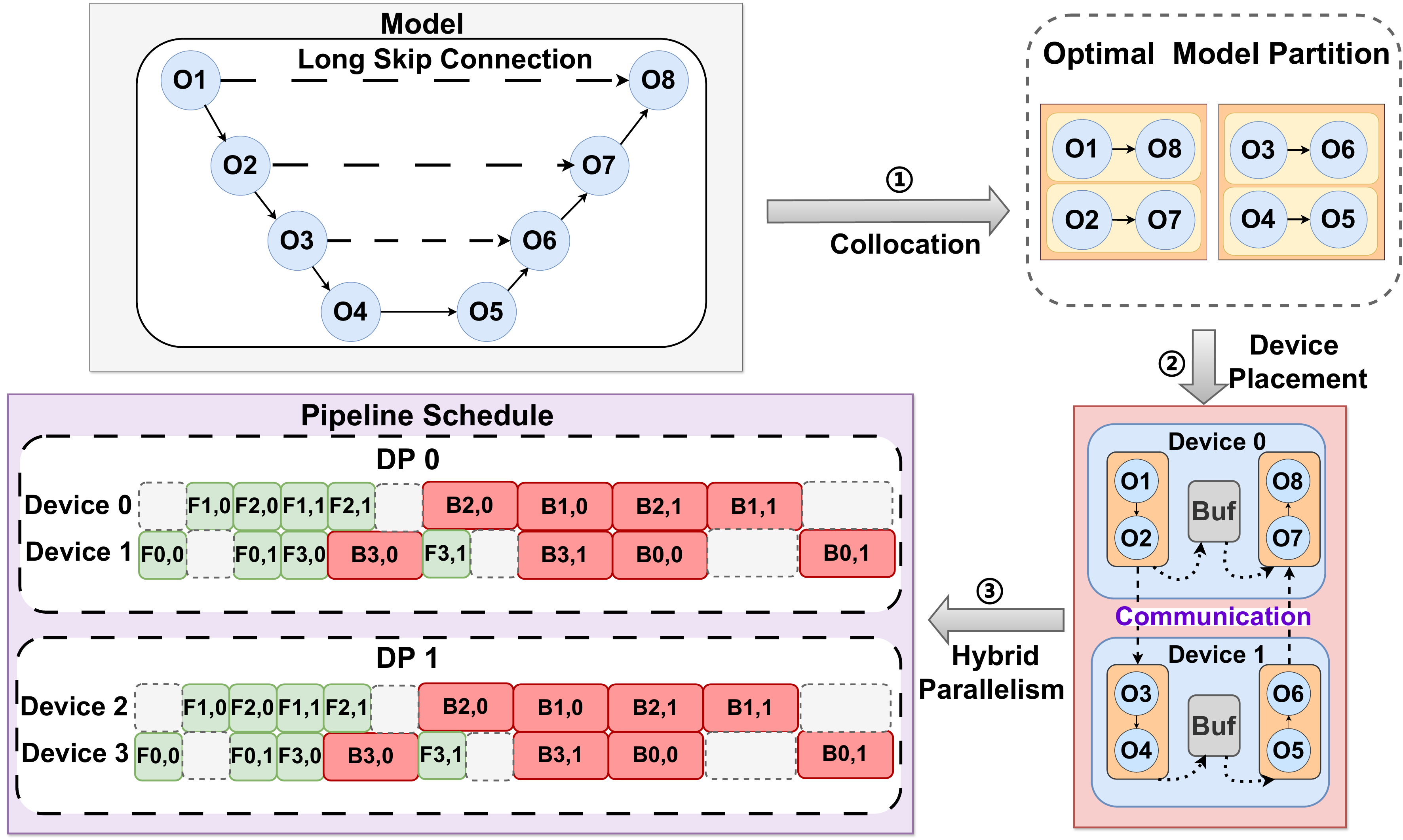}
  \caption{System overview. Our model-aware auto-parallelism framework consists of (1) skip-aware model partitioning via dynamic programming, (2) a unified pipeline schedule synthesizer that respects collocation and device constraints, and (3) a hybrid parallelism tuner that selects optimal pipeline and data-parallel configurations under memory and runtime constraints.}
  \label{fig:overview}
\end{figure}

In this work, we present \name, an automatic pipeline parallelism system designed for efficiently training diffusion models. 
Figure~\ref{fig:overview} illustrates an overview of \name, following our key observation that communication induced by skip connections can be eliminated by collocating skip-connected layers on the same device. Specifically, we allocate local buffers for skip activations and their gradients. Skip activations are stored locally during the forward pass, and these values are retrieved for gradient computation in the backward pass. This mechanism eliminates inter-device skip transfers while preserving algorithmic benefits of using skip connections. \name\ is composed of three components. 

\noindent\textbf{Model Partitioning with Collocation Constraints.}  
We decompose the model into fine-grained operations and apply a dynamic programming algorithm to assign operations to pipeline stages. Each stage must satisfy skip-connection colocation constraints, and the objective is to minimize the maximum execution time across all stages. To do so, we profile layer runtimes and solve a runtime-aware partitioning problem.

\noindent\textbf{Pipeline Scheduling with Skip-Aware Placement.}  
Given the partitioned model, we formulate a scheduling problem that assigns stage–microbatch pairs to devices and time steps. The scheduler enforces execution order, device exclusivity, and collocation constraints. For example, when the number of pipeline stages equals the number of devices, the schedule recovers the classic 1F1B pattern; when stages are doubled to enforce symmetric colocation, it converges to the Hanayo-style wave schedule~\cite{liu2023hanayo}. Our formulation generalizes to arbitrary stage–device mappings and naturally adapts to model structural heterogenity.

\noindent\textbf{Hybrid Parallelism Optimization.}  
To scale across multiple nodes, we integrate data parallelism. We model GPU memory usage—including parameters, activations, and outputs, and compute iteration time as a function of both computation and AllReduce communication. By varying the number of pipeline stages \( P \), data-parallel replicas \( G \) and micro batch size \(b\), we search for the configuration that maximizes training throughput under hardware memory constraints.

For clarity, Table~\ref{tab:notation} summarizes the notation used throughout the paper.
\begin{table}[t]
\centering
\small
\caption{Summary of Notation}
\label{tab:notation}
\begin{tabular}{l l}
\toprule
Symbol & Description \\
\midrule
$D$ & Number of devices \\
$K$ & Number of model blocks \\
$M$ & Number of microbatches \\
$\mathcal{C}$ & Set of collocated stage pairs induced by skip connections \\
$b$ & Microbatch size \\
$P$ & Pipeline parallelism degree \\
$G$ & Data parallelism degree \\
$S_i$ & The i-th PP stage \\
$\mathcal{M}_\theta$ & Size of model parameters (bytes) \\
$\mathcal{M}_a$ & Size of model activations (bytes) \\
$\mathcal{M}_o$ & Size of stage output tensor (bytes) \\
$B_{inter}$ & Effective inter-node bandwidth \\
$B_{intra}$ & Effective intra-node bandwidth \\
$t_{lat}$ & Static latency of communication kernel \\
$t_f^l$ & Forward time of layer $l$ \\
$T_f$ & Forward execution time of a stage \\
$T_{sched}$ & Total pipeline schedule time per iteration \\
$\mathcal{M}_{peak}$ & Peak memory consumption \\
\bottomrule
\end{tabular}
\end{table}

\section{Model Partitioning with Skip-Aware Constraints}
\label{sec:partition}

\chen{compress to less than 2 page columns}

\subsection{Computation Imbalance and Constraints}

\chen{compress} 
Effective model partitioning is essential for achieving high throughput in pipeline-parallel training~\cite{huang2019gpipe}. 
For a synchronized pipeline parallelism algorithm with $s$ stages and $m$ microbatches, the total pipeline latency is formulated as $\sum_{i=1}^s T_i + (m-1) \cdot \max_{1\le j \le s} \{T_j\}$, where $T_i$ is the computation time of the $i$-th stage. For simplicity, we only consider forward execution time at each stage, since backward execution time is generally estimated to be $2\times$ forward time. In this setup, the training iteration time is bounded by the slowest pipeline stage. Therefore, a well-balanced partition that can evenly distribute workload across stages is desired to avoid idle time and maximize hardware utilization. 

\begin{figure}[!htbp]
  \centering
  \includegraphics[width=0.7\linewidth]{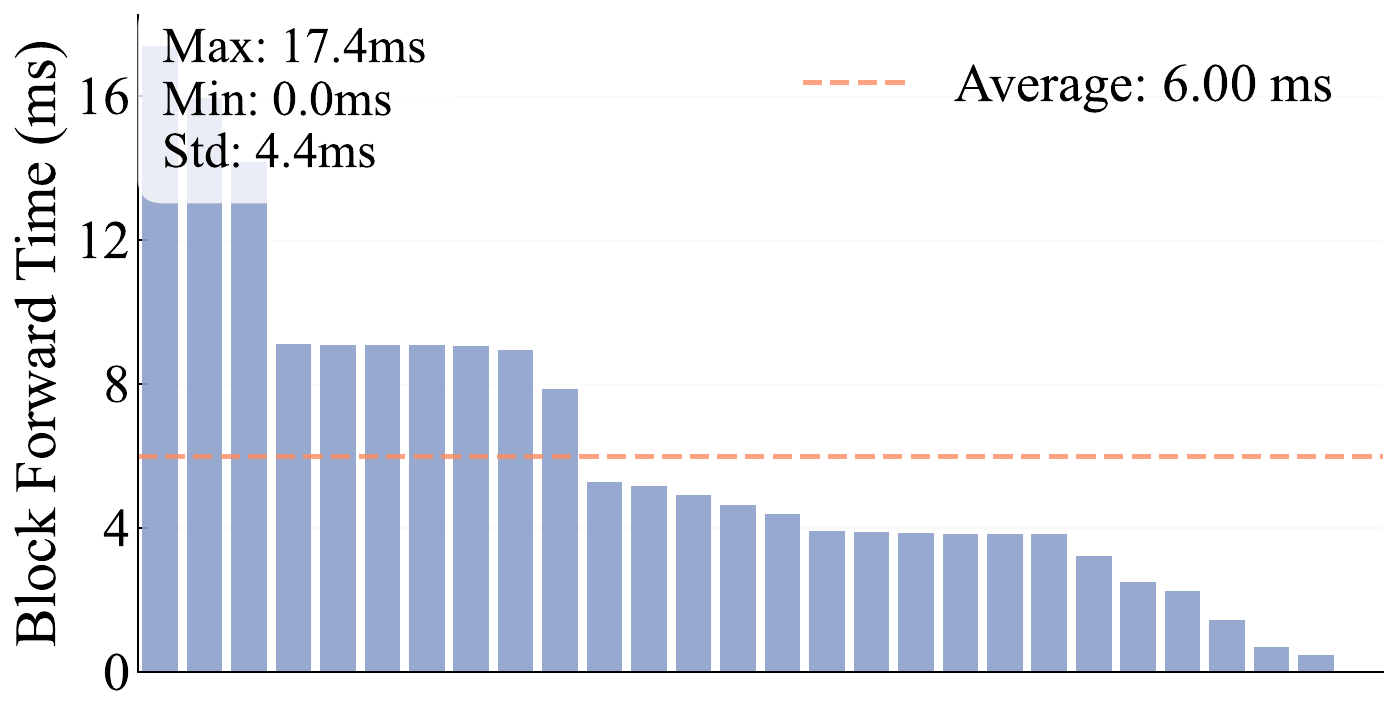}
  \caption{Per-block forward time in SDv2 model with 25 blocks, sorted descending. Heavy-tail imbalance means pipeline throughput is gated by a few slow blocks.}
  \label{fig:sd_skew_before}
\end{figure}

However, diffusion models, particularly those with UNet-style architectures, introduce additional complexity due to their encoder-decoder design and the presence of long skip connections. These inherent structural characteristics result in substantial computational heterogeneity, which is not present in conventional transformer models. As illustrated in Figure~\ref{fig:sd_skew_before}, the maximum per-stage forward time in a block-wise partitioning of SDv2 is up to \textbf{3$\times$} higher than the average, highlighting the severity of load imbalance. To tackle this problem, one common approach is to solve it with linear partition. However, skip connections complicate classical linear partitioning strategies by introducing non-local data dependencies between encoder and decoder blocks. Layers connected by these links are required to be assigned to stages that can be placed on the same device, increasing the complexity of pipeline scheduling design.

\subsection{Dynamic Programming-Based Partitioning Algorithm}

\chen{Where is your novelty? Please emphasize your contribution and compress regular techniques.}
To decouple the optimization problem including model stage partition and pipeline schedule, we simplify it by assigning symmetric stages to the same device. This aligns well with diffusion models since they also adopt a symmetric encoder-decoder architecture. We begin by factorizing the model into fine-grained operations, where each operation is treated as an atomic unit. To explicitly handle skip connections which are common in UNet-based diffusion models, we encapsulate the inputs and outputs of each operation in a dictionary structure that includes an additional field for storing skip connections. This transformation allows the model to be represented as a linear sequence of operations suitable for pipeline partitioning.

Denote the ordered sequence of operations be $\mathcal{L}=\{\ell_1,\ell_2,\ldots,\ell_{op}\}$, where \( op \) is the total number of operations. The goal is to partition \( \mathcal{L} \) into \( p \) pipeline stages, subject to skip connection constraints. Specifically, if a skip connection exists between operations $\ell_i$ and $\ell_j$ such that \( |i - j| > 1 \), and $\ell_i$ is assigned to partition \( q \), then $\ell_j$ must be assigned to partition \( p - q + 1 \). This symmetric placement ensures that skip-connected blocks reside on the same device, thereby eliminating inter-device communication for skip connections. Our goal is to find a partitioning that minimizes the maximum stage forward time across all pipeline stages—the bottleneck that determines overall throughput:
\begin{equation}
\label{formular:dp_target}
    \{ S_1, \dots, S_p \} = \arg\min \left( \max_{m \in [1, p]} (T^{S_m}_f + \lambda \frac{\mathcal{M}_a^{S_m}}{B_\text{inter}}) \right),
\end{equation}

where $T^{S_m}_f$ indicates the forward execution time of stage $m$, $\lambda$ the hyperoptable weight of activation p2p communication time, $\mathcal{M}_a^{S_m}$ the activation size calculated by stage $m$, and $B_\text{inter}$ denotes the efficient inter-node bandwidth since we consider the worst case where PP traffic is placed on the scale-out network. To solve the constrained partitioning problem, we extend the classical linear partition to a bidirectional problem. The algorithm maintains a cost table $dp(i, j, k)$ indicating optimal value of $k$ partitions over $\{\ell_1,\cdots,\ell_i\}\cup\{\ell_j,\cdots,\ell_{op}\}$. Then it can be formulated as:
\begin{equation}
    L(i', i) = \lambda \left(t_\text{lat} + \frac{\mathcal{M}_a^{S_i}}{B_\text{inter}}\right) + \sum_{i'<l\leq i}t^l_f,
\end{equation}
\begin{equation}
    R(j,j') =\lambda\left(t_\text{lat} + \frac{\mathcal{M}_a^{S_{j-1}}}{B_\text{inter}}\right) + \sum_{j\leq l< j'}t^l_f,
\end{equation}
\begin{equation}
\begin{aligned}
    dp(i,j,k) = \min_{i'<i, j'>j}\left\{\max\{dp(i',j',k - 2), L(i', i), \right. 
    \\
    \left. R(j,j'),c(i',i,j,j')\}\right\},
\end{aligned}
\end{equation}
\begin{equation}
    target=\min_{i<{op}}dp(i, i+1, p).
\end{equation}

\begin{figure}[t]
  \centering
  \begin{subfigure}[b]{0.48\linewidth}
    \centering
    \includegraphics[width=\linewidth]{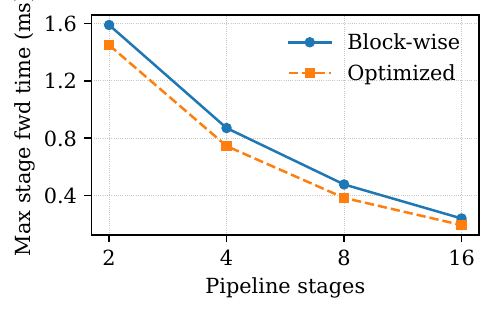}
    \caption{0.8B Hunyuan-DiT Model}
  \end{subfigure}
  \hfill
  \begin{subfigure}[b]{0.48\linewidth}
    \centering
    \includegraphics[width=\linewidth]{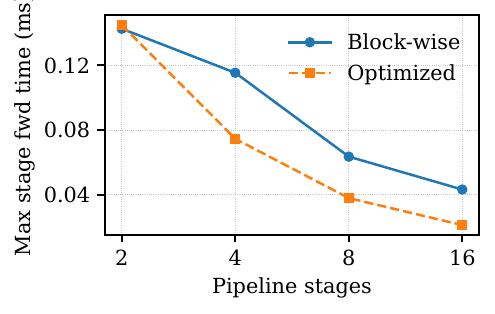}
    \caption{1.7B SDv2 Model}
  \end{subfigure}

  \caption{Performance comparison between our balanced model partition and block-wise stage assignment. Forward time is obtained by taking maximum stage time, both running with micro batch size 32. For models with imbalance computation workload e.g. SDv2, forward time can be reduced by 51.2\%.}
  \label{fig:partition_perf}
  \vspace{-0.5em}  
\end{figure}

Here, $t_f^i$ indicates the forward time of layer $i$, $t_\text{lat}$ denotes the statical latency of communication kernel, $c(i', i,j,j')$ represents the constraint penalty function within layer index $(i',i', j,j')$. The value is set to 0 and otherwise infinity when all skip connection constraints are met. Considering a satisfied constraint for example, when there exists a skip connection pair with layer index $(c_1, c_2)$, if $c_1 \leq i'$ then $c_2\geq j'$, and if $i'<c_1 \leq i$ then $j\leq c_2< j'$, otherwise the constraints are violated and $c(i',i,j,j')$ is set to infinity. After updating the dp table, we scan through and backtrack it to find the optimal partition. However, directly implementing the algorithm results in complexity of $O(pn^4)$, and is not portable for runtime deploying when the model contains hundreds of layers. To reduce the overhead in model partition, we optimize the algorithm by reusing the index, i.e. when fixing $i,i'$, $j'$ can only scan once and the complexity is reduced to $O(pn^3)$. The pseudocode is outlined in Algorithm~\ref{alg:balanced_dp_impl}. Figure~\ref{fig:partition_perf} shows the result of applying our partitioning algorithm to the same model, each running with micro batchsize 32. For Hunyuan-DiT which has similar DiT blocks, the maximum stage forward time has minor decrease, while for SDv2 which has encoder and decoder with different resolutions it reaches significant improvement. Compared to naive block-wise strategies, skip-aware balanced approach improves per-stage forward time by up to \textbf{51.2\%}, reducing pipeline bottlenecks and enabling more efficient training at scale. 

\begin{algorithm}[t]
\caption{Skip‑Aware Balanced Partitioning}
\label{alg:balanced_dp_impl}
\begin{algorithmic}[1]
\Require costs $t_f^1\dots t_f^{op}$, partitions $p$, skip constraints $c$
\State pre‑compute prefix $L[i] = \sum_{j=1}^i t_f^j$, $R[i]=\sum_{t=j}^{op}t_f^j$
\State Initialize $\text{dp}(i, j) \gets \max\{L[i], R[j], c(1,i,j,{op})\}$ 
\For{$k = 2$ \textbf{to} $p / /2$}
\For{$ l < {op}-k, l'<l$}
\State ${C_\text{enc}} \gets L[l] - L[l']$
\State $r' \gets {op}-k+1$; 
\For{$r={op}-k \text{ to } l+1$}
\While{$r' > r$ \textbf{and} $C_\text{dec}>\text{dp}(l', r')$} 
\State $C_\text{dec}\gets R[r] - R[r'] + c(l',l,r,r')$
    \State $r'\gets r'-1$ 
\EndWhile
\EndFor

\State $\text{dp}(l,r) \gets \min\bigl(\text{dp}(l,r),$ 
\State $\qquad\quad \max\left(\text{dp}(l',r'), C_{\text{enc}}, C_{\text{dec}}\right)\bigr)$

\EndFor
\EndFor
\State \Return back‑tracked cut positions ${b_m}$
\end{algorithmic}
\end{algorithm}
\vspace{-10pt}

\section{Pipeline Scheduling under Collocation Constraints}
\label{schedule}

\chen{compress to less than 1.5 page columns}

\subsection{Problem Setup}
In section~\ref{sec:partition} we optimize the scheduling step with assumption of full computation overlap, by minimizing the maximum stage execution time. However, pipe schedule always include warm up, stable and cool down stages where pipeline bubbles emerge. To further decrease the bubble ratio and find a feasible execution order that satisfies the skip connection constraint, we now consider the problem of pipeline scheduling where computation stages are assigned to resource grid composed of physical devices and time slots, under skip connection collocation constraints. Our objective is to minimize total training latency, measured by the number of scheduling steps required to complete one full pass of all microbatches through the pipeline. Here, a scheduling step represents a unit execution slot used to characterize the relative ordering and overlap of pipeline stages under collocation constraints, rather than an exact measure of wall-clock time. This abstraction allows us to reason about pipeline structure and bubble behavior, while the actual iteration time is modeled separately using profiled stage runtimes in Section~\ref{sec:hybrid}.

 For ease of representation, assume that a partitioned model with \( S  \) stages (corresponding to forward and backward passes), executed over \( M \) microbatches and \( D \) available devices.
We execute $M$ microbatches under this mapping.
We denote the set of collocated stage pairs derived from skip connections as $\mathcal{C}$. 
The scheduling horizon is slacked to \( T = S \cdot M \) steps. Let \( x_{s,m,d,t} \in \{0,1\} \) denote a binary decision variable indicating whether stage \( s \) of microbatch \( m \) is scheduled on device \( d \) at time step \( t \). We define the following auxiliary variables: \( \text{device}_s = \sum_{d,t} d \cdot x_{s,m,d,t} \) is the device to which stage \( s \) is assigned, and \( \text{time}_{s,m} = \sum_{d,t} t \cdot x_{s,m,d,t} \) is the time step at which stage \( s \) of microbatch \( m \) is executed. The scheduling must satisfy the following constraints: (1) \textit{Unique Assignment}: Each stage–microbatch pair must be scheduled exactly once; (2) \textit{Device Exclusivity}: Each device can run at most one stage per time step; (3) \textit{Fixed Device Mapping}: Each stage must be consistently mapped to a single device; (4) \textit{Collocation Constraints}: Skip-connected stage pairs must be placed on the same device; (5) \textit{Sequential Execution}: Stages within the same microbatch must respect execution order; (6) \textit{Monotonic Microbatch Ordering}: A given stage must process later microbatches no earlier than previous ones. These constraints are fomulated by the following~\ref{form:sche_start}-\ref{form:sche_end}:
\begin{equation}
\label{form:sche_start}
\sum_{d=0}^{D-1} \sum_{t=0}^{T-1} x_{s,m,d,t} = 1, \quad \forall s, m.    
\end{equation}
\begin{equation}
\sum_{s=0}^{S-1} \sum_{m=0}^{M-1} x_{s,m,d,t} \leq 1, \quad \forall d, t.
\end{equation}
\begin{equation}
\text{device}_s = \sum_{d=0}^{D-1} d \sum_{t=0}^{T-1} x_{s,m,d,t}, \quad \forall s, m.
\end{equation}
\begin{equation}
\text{device}_{s_1} = \text{device}_{s_2}, \quad \forall (s_1, s_2) \in \mathcal{C},
\end{equation}
\begin{equation}
\text{time}_{s+1,m} \geq \text{time}_{s,m} + 1, \quad \forall s < S{-}1, m.
\end{equation}
\begin{equation}
\label{form:sche_end}
\text{time}_{s,m+1} \geq \text{time}_{s,m}, \quad \forall s, m < M{-}1.
\end{equation}
We aim to minimize the maximum finishing time across all stages and microbatches. Let \( T_{\max} \) be a slack variable satisfying:
\begin{equation}
    T_{\max} \geq \text{time}_{s,m}, \quad \forall s, m,
\end{equation}
then the primary objective is $\min T_{\max}$. For determinism and alignment, the first stage is anchored to device 0, also as a secondary heuristic, we minimize a weighted sum over device indices to improve locality and readability in the final schedule:
\begin{equation}
\min \sum_{s=0}^{S-1} (-s \cdot \text{device}_s).    
\end{equation}

\subsection{Schedule Characterization}

\begin{figure}
    \centering
    \includegraphics[width=\linewidth]{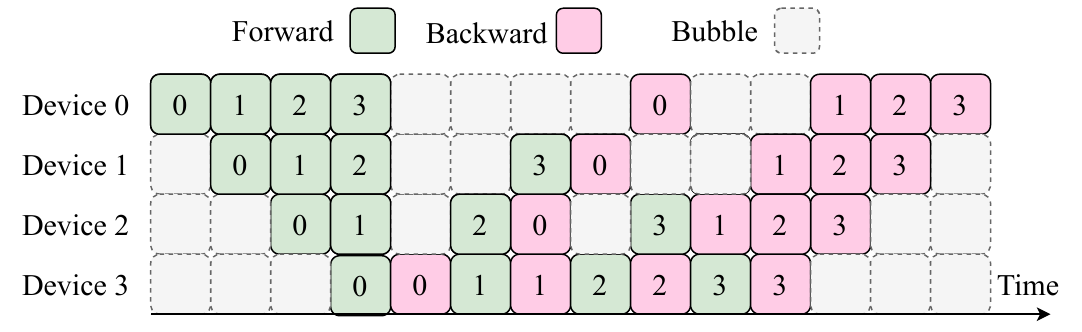}
    \caption{1F1B schedule with 4 devices, 4 pipeline stages, and 4 microbatches. The horizontal axis represents scheduling steps.}
    \label{fig:1f1b}
\end{figure}

\begin{figure}
    \centering
    \includegraphics[width=\linewidth]{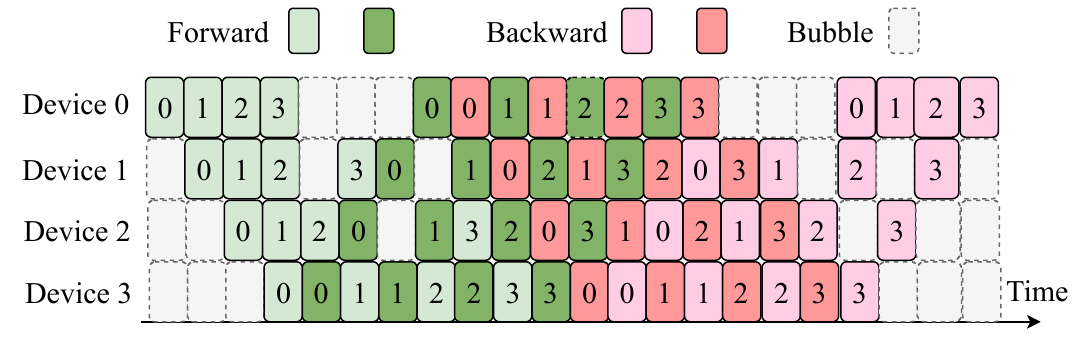}
    \caption{Wave schedule with 4 devices, 8 stages, and 4 microbatches. Symmetric layers (e.g., encoder–decoder pairs) are colocated to minimize communication of skip connections, resulting in a wave-like execution pattern. The horizontal axis represents scheduling steps. }
    \label{fig:hanayo}
\end{figure}

By solving the scheduling formulation, the execution order can be determined by $S,D$ and $\mathcal{C}$. When the number of pipeline stages equals the number of devices ($S=D$), our scheduler recovers the classic 1F1B pattern (Figure~\ref{fig:1f1b}). Under skip-connection constraints, our scheduler generates a wave-like execution pattern (Figure~\ref{fig:hanayo}). 
This schedule performs the whole forward and backward pass across different stages in a wave-like pattern. 


To reduce inter-device communication, the number of pipeline stages should ideally be minimized. The skip constrains leads to a structural lower bound on the number of pipeline stages. Following the notation in Sec.~\ref{sec:comm_model}, we use $K$ to denote the number of blocks, $D$ the number of devices,
and $a$ the skip activation size in bytes.
To colocate all symmetric encoder--decoder components on $D$ devices, we require at least $S=2D$ pipeline stages.
This structural constraint eliminates inter-device transfers of skip activations and their gradients, reducing the total communication volume
from $\left(\frac{(K+4)D}{4}-1\right)a$ to $2(D-1)a$.

Given this heuristic, we adopt the minimal-stage configuration $S=2D$ in our default scheduling strategy. This setup allows us to implement a wave schedule, where symmetric layers are mirrored across devices and executed in a staggered pattern. This execution model not only satisfies the skip connection constraints but also provides excellent overlap between forward and backward passes while minimizing idle device time. 

Although the scheduling formulation is an integer linear programming (ILP) problem and NP-hard in general, our goal is not to solve the scheduling problem at full scale, but to identify efficient execution patterns that generalize across pipeline depths and device counts. This design is motivated by the observation of pipeline parallel execution that once the pipeline reaches steady state, high-performance schedules tend to exhibit repetitive and scalable structures. Prior work~\cite{lin2024tessel} has shown that such steady-state patterns can be discovered using small-scale instances and extended to larger configurations without changing their structure.

Following this principle, we solve the ILP on a small-scale configuration (4 devices) to identify the wave scheduling pattern induced by skip-connection collocation, with a solving time on the order of several minutes depending on the number
of stages and microbatches. The resulting schedule is treated as a static template and replicated across devices for large-scale deployment. This scheduling optimization is performed offline during system initialization and does not incur runtime overhead.

We note that our wave schedule is conceptually similar to the Hanayo schedule~\cite{liu2023hanayo}. However, Hanayo is not designed to support diffusion models with skip connections, limits the number of micro-batches to the pipeline depth, and can be considered a special case of our approach.


\section{Hybrid Parallelism Strategy}
\label{sec:hybrid}
In large-scale cluster environments, heterogeneous network conditions and varying hardware capabilities across nodes pose significant challenges in determining the optimal parallel execution configuration~\cite{zheng2022alpa}. 
An appropriate parallel configuration can significantly improve resource utilization and reduce overall training time~\cite{fan2021dapple}. Leveraging the characteristics of PULSE scheduling, we model both memory consumption and computation time. By combining the strengths of pipeline parallelism (PP) and data parallelism (DP), our approach identifies an optimal parallel configuration that maximizes throughput while avoiding out-of-memory (OOM) errors.



Denote PP degree as $P$, \name\ contains $2P$ pipeline stages, and reaches the highest activation memory usage on the $P$ and $P+1$-th collocated stages since activations are retained until all micro-batches complete forward pass. For fp16 training, the peak memory $\mathcal{M}_\text{peak}$ can be formalized as:
\begin{equation}
\mathcal{M}_\text{peak} = 7 (\mathcal{M}_\theta^P+\mathcal{M}_\theta^{P+1}) +  P (\mathcal{M}_{a}^P+\mathcal{M}_a^{P+1}) \cdot b + P\mathcal{M}_o^{P-1}.
\end{equation}

Where \( b \) is the micro-batch size and $\mathcal{M}_\theta^I, \mathcal{M}_a^I, \mathcal{M}_o^I$ denote the total memory of parameters, activation, and output tensor on stages $i\in I$ obtained through profiling. Since the computation in pipeline schedule is synchronized by p2p communication, one pipeline step is gated by the slowest stage. Also, backward passes are empirically $\approx\!2\times$ the forward time, hence the total schedule time \( T_\text{sched} \) can be formalized as:
\begin{equation}
\begin{aligned}
    T_\text{sched} &= (10P-4)T_f(b) + (10P-12)\left(t_\text{lat} + \frac{b\mathcal{M}_o}{B_\text{inter}}\right) + T_\text{AR},\\
    T_f(b) &= \max_{s\in[1,2P]} T_f^s(b), \quad \mathcal{M}_o=\max_{s\in[1,2P]}\mathcal{M}_o^s.
\end{aligned}
\end{equation}
Where $T_f^s(b)$ denotes the forward time of stage $s$, the second term is the p2p communication time that increases with $P$, and $T_\text{AR}$ denotes the gradient all-reduce time for DP. Let $G$ denote the number of DP replicas (total device number $N=PG$). 
Given that DP incurs a much larger communication volume than PP~\cite{narayanan2021efficient}, we prioritize DP over PP with higher intra-node bandwidth. 
Assuming a ring-based all-reduce implementation, the communication overhead for gradient synchronization \(T_\text{AR}\) can be modeled~\cite{narayanan2021efficient} as follows:

\begin{equation}
T_\text{AR}=t_\text{lat} + \frac{2(G-1)\mathcal{M}_\theta^\text{max} }{GB_\text{intra}}, \mathcal{M}_\theta=\max_{i\in[1,2P]}\mathcal{M}_\theta^i.
\end{equation}

Our goal is to minimize the average time per training sample \(T_\text{sample}\) with device memory constraint \( \mathcal{M}_\text{peak} < \mathcal{M}_\text{limit} \), where $\mathcal{M}_{\text{peak}}$ denotes the peak per-device memory footprint during training, and
$\mathcal{M}_{\text{limit}}$ is the available per-device memory budget:
\begin{equation}
    \min_{P, G, b} T_\text{sample} = \frac{T_\text{sched}}{b \cdot P\cdot G}.
\end{equation}

We enumerate all valid factorizations of the total number of devices \( N = P \cdot G \), and for each configuration, determine the maximum feasible microbatch size \( b \in \mathcal{B}=\{1, 2, 4, \ldots\} \) that fits within memory. We then evaluate the per-sample training time and select the configuration that yields the best throughput.

\section{Evaluation}
In this section, we evaluate the effectiveness of \text{\name} for training large diffusion models. We focus on analyzing training throughput, communication volume, and scalability across a variety of model architectures by comparing our approach to several baseline methods. 

\phm{Hardware platforms.}
Experiments are conducted on two hardware configurations:
\begin{itemize}[left=0em, itemsep=2pt, topsep=2pt]
    \item $2\times$ NVIDIA V100 nodes:  Each node contains 8 V100 GPUs with 32GB memory, connected via NVLink with an intra-node bandwidth of 300GB/s. Nodes are connected through Infiniband with a bandwidth of 10GB/s.
    \item $8\times$ Ascend 910A nodes: Each node contains 8 Ascend 910A NPUs with 32GB memory. The intra-node bandwidth is 30GB/s, and the inter-node bandwidth is 19GB/s. This setup reflects a scenario with limited bandwidth.
\end{itemize}

\phm{Models.}
We train the following models: UViT~\cite{bao2023all}, Stable Diffusion v2 (SDv2)~\cite{rombach2022high, ronneberger2015u} and Hunyuan-DiT~\cite{li2024hunyuan}. In addition to their original implementation, we also scale the models by adding blocks or increasing the hidden size.
\chen{compress model introduction}

We assume that the input images have already been downsampled and processed into latent representations, and the text inputs have been processed into encoder hidden states embeddings. These preprocessing stages, including image encoding and text embedding, are assumed to be done prior to training and thus do not contribute to the throughput evaluated in the experiments. Detailed configurations for the input of models are listed below in Table \ref{tab:model_config}.
\begin{table}[!th]  
	\caption{Model Input Configuration Details } \centering
	\vspace{-2mm}	
	\label{tab:model_config}
	\begin{tabular}{lllll}  	
		\toprule   
		\textbf{Model} & \textbf{Latent Shape} & \textbf{Condition Input} \\  
		\midrule   
		\textbf{UViT} & 32x32x3 & Class Condition \\
         \textbf{SDv2} & 32x32x4 & Clip Embeddings  \\
         \textbf{Hunyuan-DiT} &64x64x4 &   Clip \& T5 Embeddings   \\
		\bottomrule    
	\end{tabular}
\end{table}

\noindent\textbf{Baselines.} We compare our method with three baselines: Hanayo~\cite{liu2023hanayo}, Megatron 1F1B~\cite{fan2021dapple} and DeepSpeed ZeRO Stage2 (ZeRO-2)~\cite{rajbhandari2020zero}. For Hanayo and 1F1B, we partition the model in a block-wise manner, assigning the encoder and decoder blocks to devices sequentially. Following the original implementation, skip connections are computed during the forward pass and stacked, transferred, and then popped out at the corresponding decoder stage. For a fair comparison, we adopt the same hybrid parallelism settings for Hanayo and 1F1B, and the same microbatch size for all baselines.

\subsection{Overall Performance}
\label{sec:main_exp}
\begin{figure*}[htbp]
\centering
\begin{subfigure}{0.49\textwidth}
    \includegraphics[width=\textwidth]{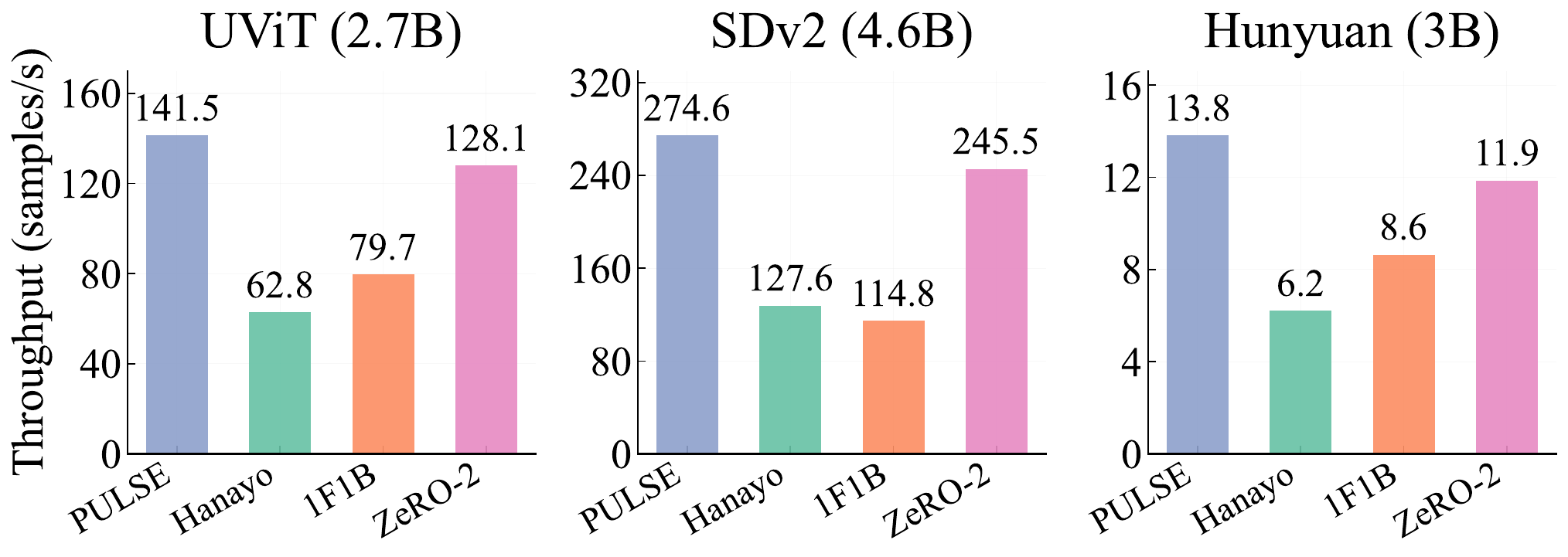}
    \caption{Throughput on a 2-node V100 Cluster}
\end{subfigure}
\hfill
\begin{subfigure}{0.49\textwidth}
    \includegraphics[width=\textwidth]{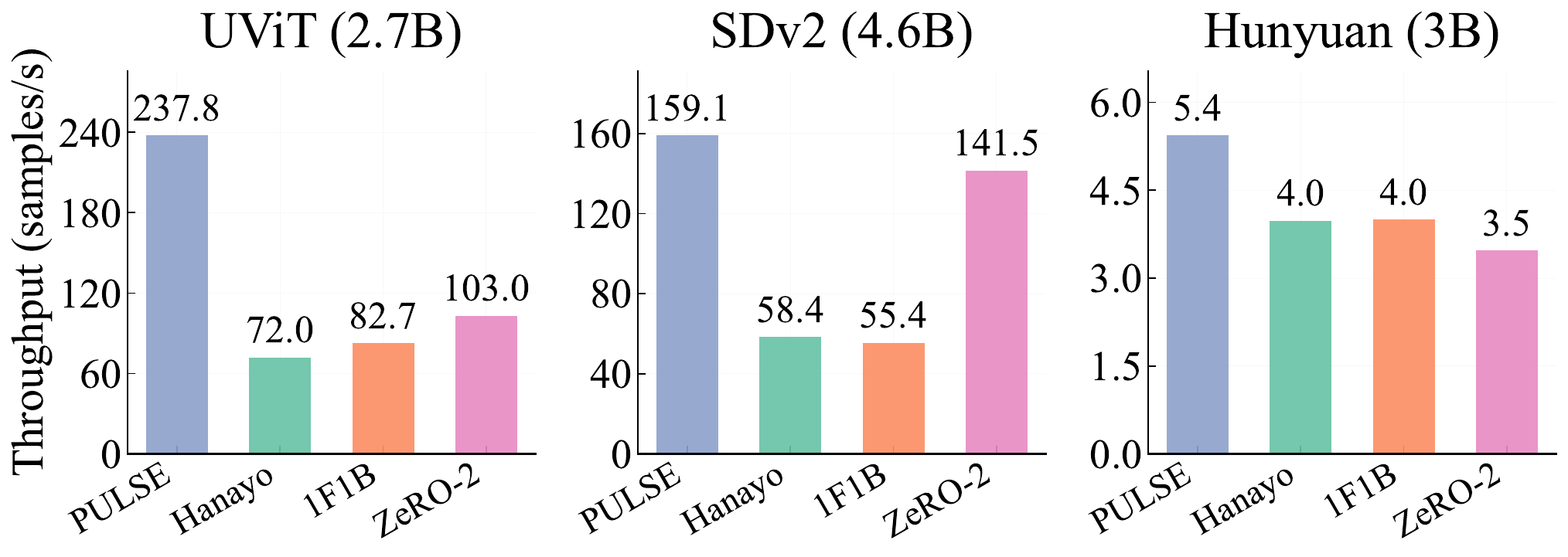}
    \caption{Throughput on a 8-node Ascend cluster}
\end{subfigure}
\caption{Evaluation Results on 2-node V100 cluster and 8-node Ascend cluster. Results show that \text{\name} can reaches up to $2.3\times$ improvement in throughput compared with PP baselines, and $1.3\times$ improvement with ZeRO-2 baseline. }
\label{fig:main_exp}
\end{figure*}    

\begin{table}[!htbp]
    \centering
    \caption{Per-sample Communication Volume (MB) on 2-node V100 cluster and 8-node Ascend 910A cluster}
    \label{tab:comm_volume_comparison}
    \begin{tabular}{lcccccc}
        \toprule
        \multirow{2}{*}{Method} & \multicolumn{2}{c}{UViT (2.7B)} & \multicolumn{2}{c}{SDv2 (4.6B)} & \multicolumn{2}{c}{Hunyuan-DiT (3B)} \\
        \cmidrule(lr){2-3} \cmidrule(lr){4-5} \cmidrule(lr){6-7}
         & V100 & 910A & V100 & 910A & V100 & 910A \\
        \midrule
        \name   & \textbf{24.19}  & \textbf{28.22}  & \textbf{6.71}   & \textbf{9.28}   & \textbf{95.29}  & \textbf{95.29} \\
        Hanayo  & 294.28 & 294.28 & 117.56 & 117.56 & 2885.57 & 4385.47 \\
        1F1B    & 102.80 & 403.13 & 61.05  & 58.97  & 916.64 & 916.64 \\
        ZeRO-2  & 318.08 & 248.59 & 276.48 & 216.15 & 2981.18 & 4385.47 \\
        \bottomrule
    \end{tabular}
\end{table}

\begin{figure}[!htbp]
    \centering
    \includegraphics[width=\linewidth]{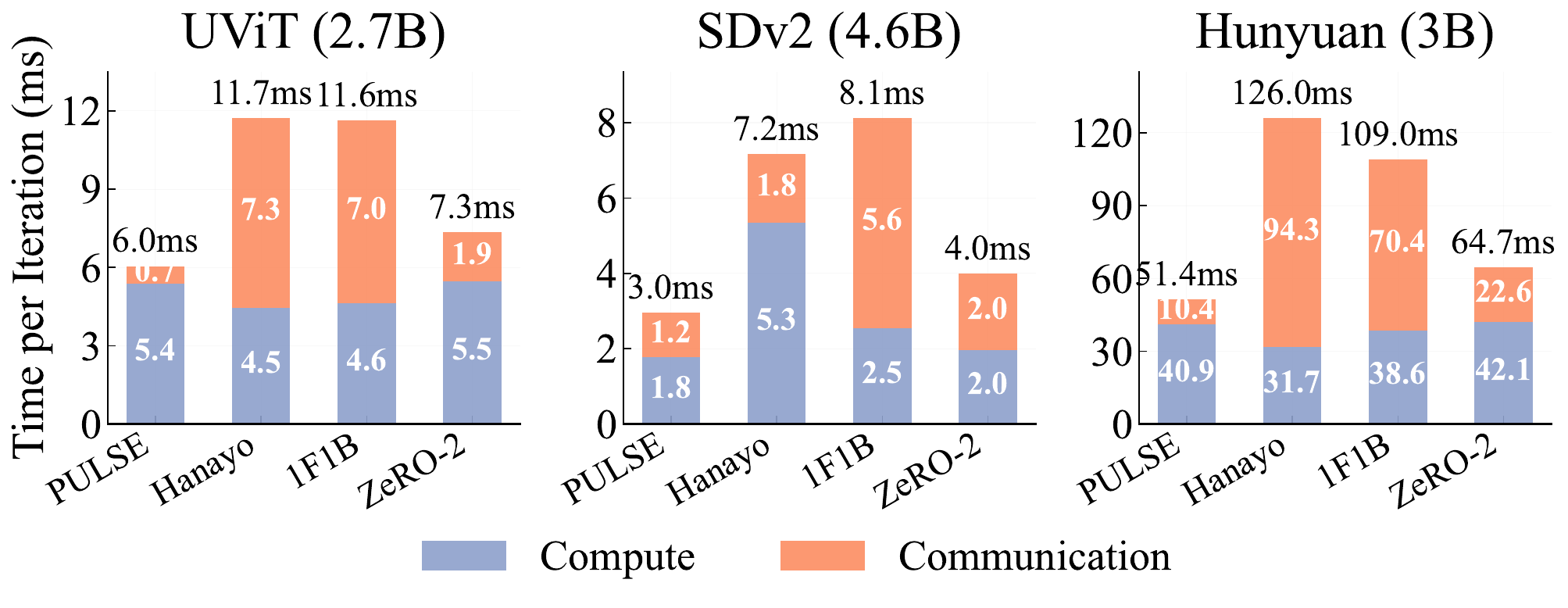}
    \caption{Per‑sample time breakdown on a 2‑node V100 cluster. \text{\name} eliminates most skip traffic and thus spends $\le 30\%$ of its iteration on communication.}
    \label{fig:time_breakdown}
\end{figure}

Figure~\ref{fig:main_exp} and table~\ref{tab:comm_volume_comparison} present training throughput and communication volume over different models and cluster settings. On a 2-node V100 cluster, \text{\name} achieves 274.6 samples per second training throughput, speeding up 140\% over Megatron 1F1B for a SDv2 (4.6B) model. \text{\name} also outperforms the strong data parallel ZeRO-2 baseline with around 10\% improvement, because in this setting, model size is relatively large. The reduce-scatter communication for gradient and optimizer state takes dominance over the p2p communication for model activations in pipeline parallelism, resulting in overhead larger than the pipeline bubble rate in \name. To illustrate this, Figure~\ref{fig:time_breakdown} shows the time breakdown per sample on 2 V100 nodes. The computation time is obtained by adding the forward time and the backward time, and then averaged by a global batchsize. For communication time, \text{\name} and Megatron 1F1B mainly involve synchronized p2p communication, so we calculate it by $\max(t_{\text{send\_a}},t_{\text{recv\_a}})+\max(t_{\text{send\_g}},t_{\text{recv\_g}})$. For DeepSpeed ZeRO-2, we calculate by counting all the reduce-scatter time. The result shows that the computation time is similar, while \text{\name} reduces the communication time by up to 90\% compared with Megatron 1F1B, and 63\% compared with ZeRO-2. 

On 8 Ascend 910A nodes, \text{\name} reaches $2.87\times$ acceleration over Megatron 1F1B and $2.31\times$ over DeepSpeed ZeRO-2. These results validate the effectiveness of our skip-aware partitioning and communication-optimized scheduling in large-scale, multi-node environments. Even in low-batch, high-resolution settings (e.g., Hunyuan-DiT), our method maintains high throughput while reducing cross-device communication, demonstrating its practical scalability and generality.

\subsection{Model Scalability Evaluation}
\label{sec:scalability}
\begin{figure}[!htbp]
  \centering
  \includegraphics[width=\linewidth]{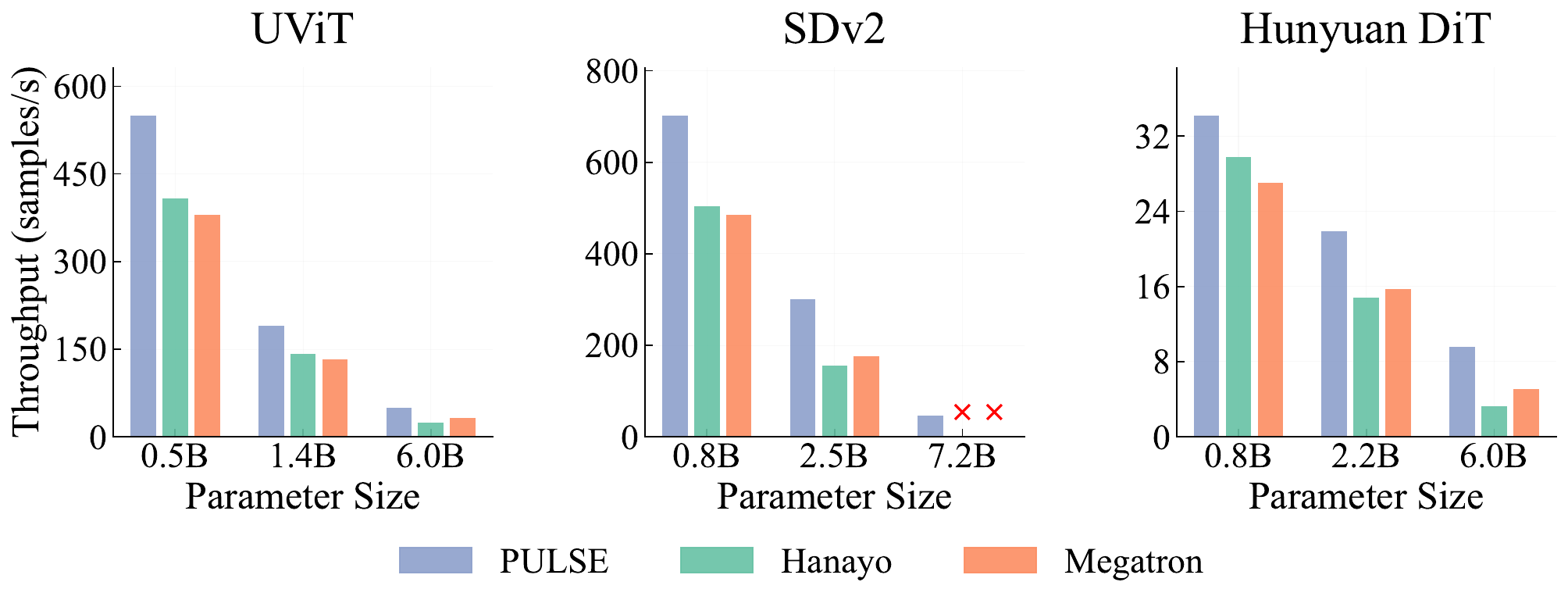}
  \caption{Training throughput (samples/sec) across different model sizes and architectures. Our method consistently outperforms most baselines, especially on large models where pipeline depth and skip communication become critical.}
  \label{fig:scalability_throughput}
\end{figure}

We evaluate how \text{\name} improves diffusion‑model training when scaling the model parameters. Experiments are run on 2 V100 nodes using three models at multiple parameter scales. The baselines are Hanayo and Megatron 1F1B. Figure~\ref{fig:scalability_throughput} shows the throughput versus model size. \text{\name} outperforms baselines over all model sizes, and the improvement is obvious with larger model sizes. For example, \name\ reaches a 57.1\% improvement in throughput when increasing the UVIT parameters to 6.0B, and enables training of a 7.2B SDv2 while other methods are out of memory. These results substantiate our core claim: skip‑aware collocation markedly reduces communication overhead, thereby boosting throughput and enabling the training of larger diffusion models on fixed hardware budgets.

\subsection{Ablation Study}

\subsubsection{Ablation on Skip-Aware Dynamic Partitioning}
\label{sec:ablation_partition}

We evaluate the effect of our dynamic programming-based skip-aware partitioning strategy by comparing it against a conventional block-wise partition baseline. The baseline assigns consecutive encoder and decoder blocks to pipeline stages without accounting for intra-block heterogeneity. All other experimental settings (e.g., hybrid parallel configuration and microbatch size) are kept consistent with those used in the model scalability experiments.

\begin{figure}[h]
  \centering
  \includegraphics[width=\linewidth]{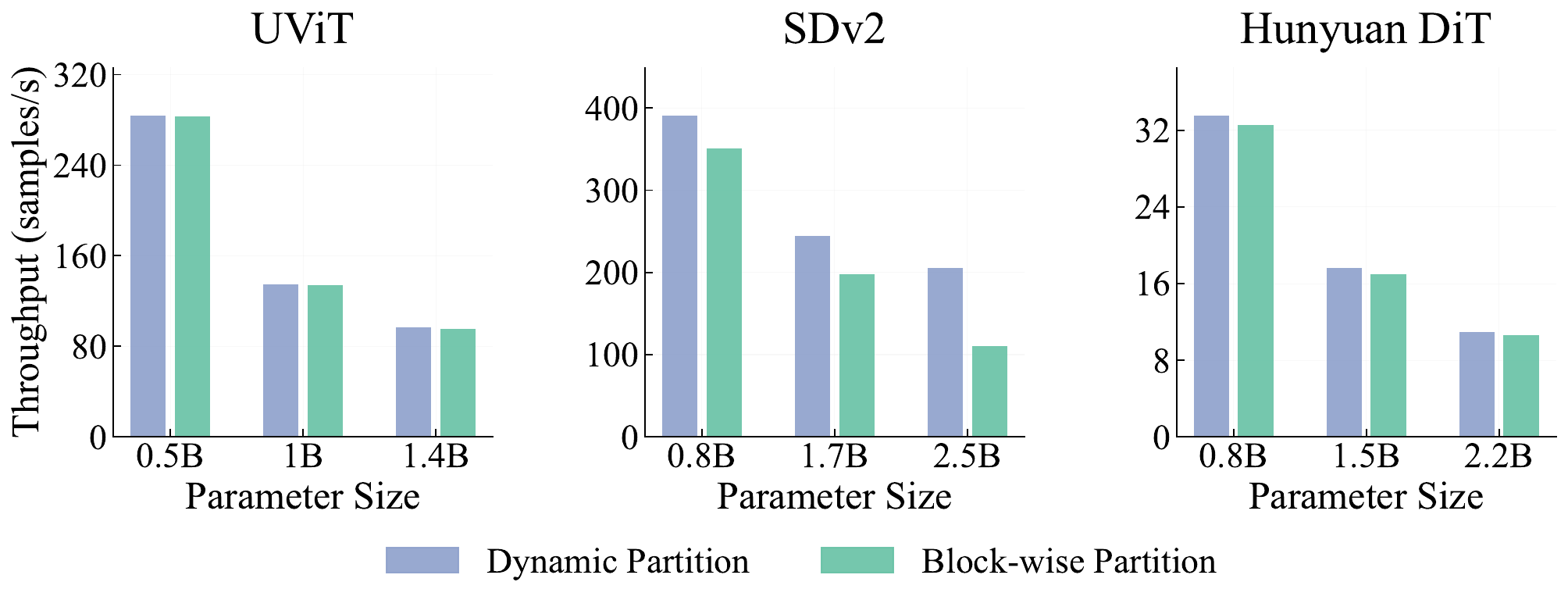}
  \caption{Ablation study on model partitioning. The improvement is most significant on SDv2.}
  \label{fig:ablation_partition}
\end{figure}

Figure~\ref{fig:ablation_partition} presents the throughput comparison across three models and parameter scales. Our method achieves significant performance improvements on SDv2, with up to 85.5\% higher throughput (2.5B model), due to its highly imbalanced encoder-decoder structure with downsampling and upsampling stages. For UViT and Hunyuan-DiT, the throughput gains are marginal (1–2\%), as these models consist of uniformly structured transformer blocks with relatively balanced computation. In such cases, block-wise partitioning introduces less load imbalance, and the benefit of dynamic partitioning is less pronounced.

\subsubsection{Ablation on Hybrid Parallelism}
\label{sec:ablation_hybrid}

\begin{figure}[!htbp]
  \centering
  \includegraphics[width=\linewidth]{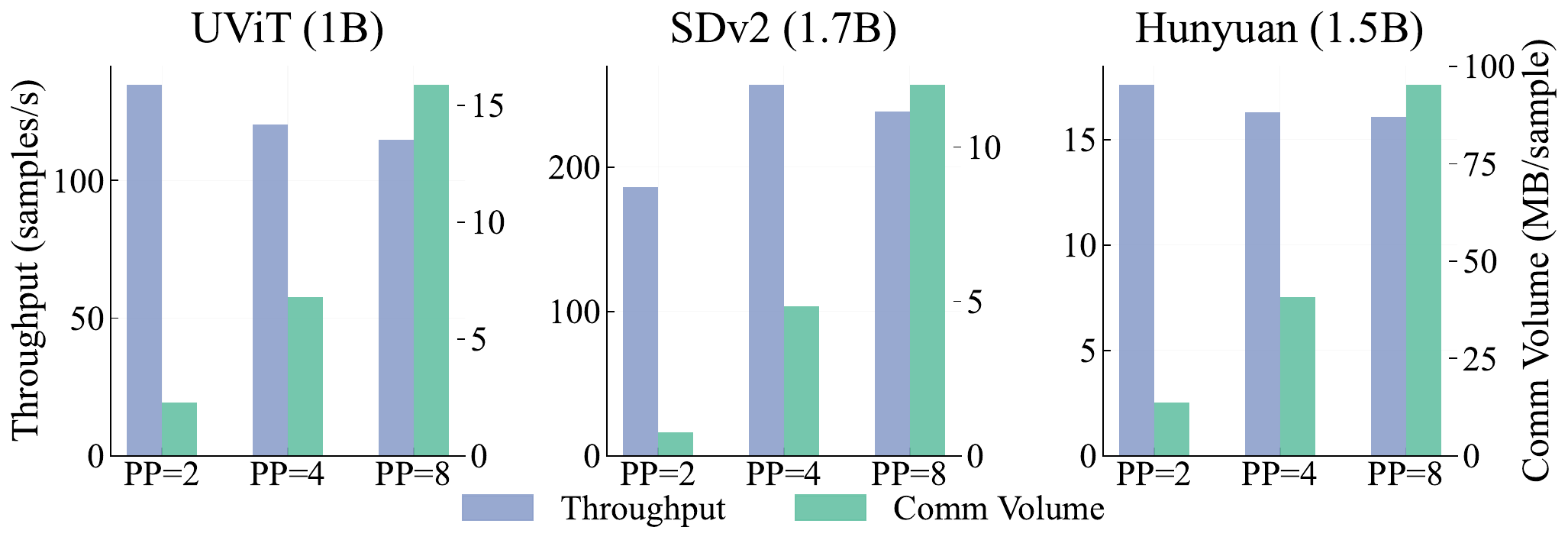}
  \caption{
  Ablation study on hybrid parallelism configurations.
  }
  \label{fig:HybridAblation}
\end{figure}

We conduct an ablation study on hybrid parallelism configurations to understand how different hybrid parallelism configurations affect training performance. The experiment is performed on 8 NVIDIA V100 GPUs by varying the pipeline parallelism degree \( P \in \{2, 4, 8\} \) and adjusting data parallelism accordingly (\( G = 8/P \)). Figure~\ref{fig:HybridAblation} summarizes both training throughput and point-to-point communication volume per sample across three models: UViT (1B), SDv2 (1.7B), and Hunyuan-DiT (1.5B). For UViT and Hunyuan-DiT, throughput decreases monotonically as the pipeline parallelism degree increases (e.g., UViT drops from 134.7 to 114.7 samples/sec). These transformer-based models feature uniform computation profiles across layers, and thus derive no significant benefit from deeper pipeline partitioning. Increasing \( P \) only exacerbates communication overhead and leads to diminishing returns. SDv2 (1.7B), in contrast, benefits from moderate pipeline depth. With \( P=2 \), the model is constrained by memory, requiring a microbatch size of only 16. With \( P=4 \), the reduced per-stage memory footprint permits a larger microbatch size of 32, resulting in a significant throughput boost from 186.2 to 257.0 samples/sec. However, pushing further to \( P=8 \) degrades performance due to excessive inter-stage communication. Communication volume increases roughly linearly with the number of pipeline stages across all models. For instance, SDv2's per-sample communication volume rises from 0.77MB at \( P=2 \) to 4.84MB at \( P=8 \), while Hunyuan-DiT sees a jump from 13.6MB to 95.3MB. 

\section{Discussions}
\subsection{Scope of Pipeline Optimization}

This work focuses on optimizing pipeline parallelism for diffusion models with long skip connections, and targets communication inefficiencies that arise at the granularity of pipeline stages. Accordingly, we do not explicitly incorporate tensor parallelism (TP) into the optimization space. TP primarily partitions computation within individual layers, such as attention or feed-forward blocks, and mainly affects intra-layer compute
parallelism and collective communication patterns (e.g., AllReduce or AllGather). In contrast, \name\ addresses a different performance bottleneck: excessive point-to-point activation communication induced by long skip connections in UNet-style architectures, which dominates inter-stage traffic in naive pipeline parallelism. Our core design operates at the level of pipeline stages and layer placement. The communication optimized by \name\ is therefore structurally orthogonal to the collective communication introduced by TP, and incorporating TP does not alter the observation that skip-connected encoder--decoder layers benefit from being colocated.

At the same time, TP can be naturally composed with \name. Each pipeline stage generated by \name\ can internally employ TP to further parallelize computation-heavy layers, following common practice in large-scale training systems. Such composition does not invalidate our conclusions on skip-aware pipeline optimization. We leave a systematic exploration of the combined TP-PP design space to future work.

\subsection{System Applicability}

\name\ is primarily designed for encoder-decoder architectures with structured skip symmetry, as commonly found in diffusion models. These models exhibit large, high-resolution skip activations that make skip-induced communication a dominant performance bottleneck, rendering skip-aware collocation particularly effective.

For architectures with irregular or partial skip patterns, skip-aware collocation can still be applied selectively to high-volume skip pairs, while other connections are routed via runtime communication. Extending \name\ to automatically identify and optimize such partial collocation opportunities, as well as to support more general skip
structures, is an interesting direction for future work.
\section{Related Work}
\subsection{Models with Long Skip Connections}
UNet backbones popularized by Stable Diffusion~\cite{rombach2022high} rely on symmetric encoder–decoder skip connections to preserve spatial detail during iterative denoising. Recent text‑to‑image systems have replaced most convolutions with vision transformers, e.g. DiT~\cite{peebles2023scalable}, UViT~\cite{bao2023all}, and Hunyuan-DiT~\cite{li2024hunyuan}, yet they retain the long‑range skips, since ablating them degrades FID and CLIP‑score by up to 10\%\cite{li2024hunyuan}. The same pattern extends to text‑to‑video~\cite{bao2024vidu} and multi‑modal fusion models such as
TransFusion~\cite{zhou2024transfusion}. These non‑local data dependencies break the sequential assumptions of classic pipeline parallelism, and generate large activation tensors that must be sent through many layers. Our work builds on this observation, trying to preserve the beneficial skip connections in model architecture while eliminating their cross‑device traffic to minimize training cost.

\subsection{Parallel Mechanism for Multimodal Models}
With diffusion models now being the core of many multi‑modal pipelines, 
many works have turned to optimizing and accelerating the training 
of these richer, heterogeneous architectures. Recent efforts to optimize multi-modal model training focus on heterogeneous parallelism and bubble minimization, but largely overlook the communication patterns induced by long-range skip connections.

Several general-purpose pipeline parallelism frameworks are capable of correctly handling skip connections. TorchGpipe~\cite{kim2020torchgpipe}, an early PyTorch implementation of GPipe, executes pipeline
stages within a single process and relies on remote memory access to transmit activations, including skip-connected tensors, directly to their consumer stages. Similarly, PiPPy~\cite{pippy2022}, the former default pipeline parallelism framework in PyTorch, leverages Torch FX to trace computation graphs and identify skip connections, issuing point-to-point communication on demand when skip activations are consumed. These frameworks treat skip connections as a runtime communication feature, ensuring correctness of execution while the pipeline schedule (e.g., 1F1B pipeline schedule) remains unchanged.

Beyond these frameworks, Alpa~\cite{zheng2022alpa} searches the device and operator space to compose data, tensors, and pipeline parallelism. 
DISTMM~\cite{huang2024distmm} and DistTrain~\cite{zhang2024disttrain} handle module heterogeneity in multi‑modal LLMs by assigning different DP/TP strategies per sub‑module. Spindle~\cite{wang2024efficient} decomposes multi-task multimodal models into waves and jointly optimizes the workload with submodule dependencies. DiffusionPipe~\cite{tian2024diffusionpipe} fills the non-trainable VAE encoders into pipeline bubbles, but still partitions the UNet itself sequentially and therefore incurs skip‑edge traffic. Graphpipe~\cite{jeon2024graphpipe} considers the multi-input branch of multimodal models and applies graph parallelism in standard pipeline parallelism; however, the graph partition algorithm fails on the UNet skip connection structure, which is different from the multi-branch input structure.

None of the above frameworks explicitly model the symmetry constraint imposed by encoder–decoder connections. They either treat skip communication as a runtime concern or operate at a submodule level under different backbone assumptions. We complement this line of work by adding a skip‑locality dimension to the optimization search space, bridging the gap between diffusion model architecture and distributed training efficiency.

\section{Conclusion}
We presented \name, an auto-parallelism framework for accelerating large diffusion model training with the objective of minimizing communication overhead by collocating skip-connected layers on the same device. \name\ combines an efficient constraint-based model partitioning with pipeline parallelism scheduling and hybrid parallelism optimization to address UNet architecture challenges. PULSE reduces communication volume by up to 89\% and increases throughput by $2.3\times$ compared with state-of-the-art methods, providing a feasible solution for efficiently training large diffusion models on commodity hardware.
\section{Acknowledgments}


This work was supported in part by multiple funding sources, including the National Natural Science Foundation of China (NSFC) under Grant 62432008, the Research Grants Council (RGC) of Hong Kong under Grant R6021-20, Grant T43-513/23N-2, Grant C7004-22G, Grant C1029-22G, Grant C6015-23G, Grant CRS\_HKUST601/24, Grant 16207922, Grant 16207423, and Grant 16203824, the Major Key Project of Peng Cheng Laboratory under Grants PCL2024A06 and PCL2025A10, and the Shenzhen Science and Technology Program under Grant RCJC20231211085918010.

\bibliographystyle{IEEEtran}
\bibliography{main}

\end{document}